\documentclass[twocolumn]{aastex62}
\received{---}
\revised{---}
\accepted{---}
\submitjournal{ApJ}
\shorttitle{Early Low-Mass Galaxies and Star-Cluster Candidates at $z\sim6-10$}
\shortauthors{Kikuchihara et al.}

\newcommand{\MS}{{\mathrm{M}_\odot}}
\newcommand{\ZS}{{\mathrm{Z}_\odot}}
\newcommand{\UV}{\mathrm{UV}}
\newcommand{\bnB}{\ensuremath{B_{435}}}
\newcommand{\bnV}{\ensuremath{V_{606}}}
\newcommand{\bni}{\ensuremath{i_{814}}}
\newcommand{\bnY}{\ensuremath{Y_{105}}}
\newcommand{\bnJ}{\ensuremath{J_{125}}}
\newcommand{\bnJH}{\ensuremath{J\!H_{140}}}
\newcommand{\bnH}{\ensuremath{H_{160}}}
\newcommand{\bnKs}{\ensuremath{K_s}}
\newcommand{\aper}{\text{ap}}
\newcommand{\tot}{\text{tot}}
\newcommand{\phot}{\text{phot}}
\newcommand{\median}{\text{median}}
\newcommand{\micro}{\mu\mathrm{m}}
\newcommand{\SN}{\mathrm{S}/\mathrm{N}}

\usepackage{amsmath}
\usepackage{ascmac}
\usepackage{bm}

\begin{document}


\title{Early Low-Mass Galaxies and Star-Cluster Candidates at $z\sim6-9$\\ Identified by the Gravitational Lensing Technique and Deep Optical/Near-Infrared Imaging}


\newcommand{\ICRR}{Institute for Cosmic Ray Research, The University of Tokyo, 5-1-5 Kashiwanoha, Kashiwa, Chiba 277-8582, Japan}
\newcommand{\IPMU}{Kavli Institute for the Physics and Mathematics of the universe (Kavli IPMU, WPI), The University of Tokyo, 5-1-5 Kashiwanoha, Kashiwa, Chiba 277-8583, Japan}
\newcommand{\TodaiPhys}{Department of Physics, Graduate School of Science, The University of Tokyo, 7-3-1 Hongo, Bunkyo-ku, Tokyo 113-0033, Japan}
\newcommand{\Sorbonne}{Sorbonne Universit{\'e}s, UPMC-CNRS, UMR7095, Institut d'Astrophysique de Paris, F-75014 Paris, France}

\author[0000-0003-2449-6314]{Shotaro Kikuchihara}
\affiliation{\ICRR}
\affiliation{Department of Astronomy, Graduate School of Science, The University of Tokyo, 7-3-1 Hongo, Bunkyo, Tokyo, 113-0033, Japan}
\email{skiku@icrr.u-tokyo.ac.jp}

\author[0000-0002-1049-6658]{Masami Ouchi}
\affiliation{\ICRR}
\affiliation{\IPMU}

\author[0000-0001-9011-7605]{Yoshiaki Ono}
\affiliation{\ICRR}

\author[0000-0003-4985-0201]{Ken Mawatari}
\affiliation{\ICRR}

\author[0000-0002-7636-0534]{Jacopo Chevallard}
\affiliation{\Sorbonne}

\author[0000-0002-6047-430X]{Yuichi Harikane}
\affiliation{National Astronomical Observatory of Japan, 2-21-1 Osawa, Mitaka, Tokyo 181-8588, Japan}
\affiliation{\ICRR}
\affiliation{\TodaiPhys}

\author[0000-0001-5780-1886]{Takashi Kojima}
\affiliation{\ICRR}
\affiliation{\TodaiPhys}

\author[0000-0003-3484-399X]{Masamune Oguri}
\affiliation{Research Center for the Early Universe, The University of Tokyo, 7-3-1 Hongo, Bunkyo-ku, Tokyo 113-0033, Japan}
\affiliation{\TodaiPhys}
\affiliation{\IPMU}

\author[0000-0002-6971-5755]{Gustavo Bruzual}
\affiliation{Instituto de Radioastronom\'ia y Astrof\'isica, Universidad Nacional Aut\'onoma de M\'exico, Morelia, Michoac\'an, 58089 M\'exico}

\author[0000-0003-3458-2275]{St{\'e}phane Charlot}
\affiliation{\Sorbonne}


\begin{abstract}

We present very faint dropout galaxies at $z\sim6-9$ with a stellar mass $M_\star$ down to $M_\star\sim10^6~\MS$ that are found in deep optical/near-infrared (NIR) images of the full data sets of the {\it Hubble} Frontier Fields (HFF) program in conjunction with deep ground-based and {\it Spitzer} images and gravitational lensing magnification effects.
We investigate stellar populations of the HFF dropout galaxies with the optical/NIR photometry and {\tt BEAGLE} models made of self-consistent stellar population synthesis and photoionization models, carefully including strong nebular emission impacting on the photometry.
We identify 357 galaxies with $M_\star\sim10^6-10^9~\MS$, and find that a stellar mass to UV luminosity $L_{\rm UV}$ ratio $M_\star/L_{\rm UV}$ is nearly constant at $M_\star\sim10^6-10^9~\MS$.
Our best-estimate $M_\star/L_{\rm UV}$ function is comparable to a model of star-formation duration time of 100 Myr, but $2-7$ times higher than the one of 10 Myr assumed in a previous study (at the $5\sigma$ level) that would probably underestimate $M_\star$ of faint galaxies.
We derive the galaxy stellar mass functions (GSMFs) at $z\sim6-9$ that agree with those obtained by previous studies with no $M_\star/L_{\rm UV}$ assumptions at $M_\star\gtrsim10^8~\MS$, and that extends to $M_\star\sim10^6~\MS$.
Estimating the stellar mass densities $\rho_\star$ with the GSMFs, we find that $\rho_\star$ smoothly increases from $\log(\rho_\star/[\MS~\mathrm{Mpc}^{-3}])=5.91_{-0.65}^{+0.75}$ at $z\sim9$ to $6.21_{-0.37}^{+0.39}$ at $z\sim 6-7$, which is consistent with the one estimated from star-formation rate density measurements.
In conjunction with the estimates of the galaxy effective radii $R_\mathrm{e}$ on the source plane, we have pinpointed two objects with low stellar masses ($M_\star\leq10^7~\MS$) and very compact morphologies ($R_\mathrm{e}\leq40$ physical pc) that are comparable with those of globular clusters (GCs) in the Milky Way today. 
These objects are candidates of star clusters that should be a part or a dominant component of high-redshift low-mass galaxy, some of which may be related to GCs today.

\end{abstract}




\section{Introduction} \label{sec:intro}

The stellar mass of a galaxy is an indicator of the masses that are aggregated through the previous star formation and merging processes.
Various observations and simulations have found that the stellar masses correlate with basic properties of galaxies, such as the star formation rate \citep[SFR; e.g.,][]{McLure+11, Speagle+14, Salmon+15, Santini+17, Iyer+18}, age \citep[e.g.,][]{Sparre+15}, metallicity \citep[e.g.,][]{Kojima+17, Barber+18}, and size \citep{Trujillo+04, van-der-Wel+14, Lange+15}.
These facts imply that the stellar mass plays a critical role in understanding the formation and evolution of galaxies in the early universe, which still are major open questions in astronomy today.

The number density of galaxies per stellar-mass ($M_\star$) interval, i.e., the galaxy stellar mass function (GSMF), is often used to study the evolution of the total stellar mass in the universe.
The low-mass ends of the GSMFs are especially worth investigating at high redshift, because low-mass galaxies are expected to be dominant in the early universe according to the hierarchical cosmology.
It is also worth mentioning that the low-mass galaxies ($M_\star\sim10^6-10^7~\MS$) at $z\sim6-9$ are predicted to grow up to have stellar masses comparable to that of the Milky Way at $z\sim0$ \citep[e.g.,][]{Behroozi+13, Behroozi+18}.

However, the low-mass ends of the GSMFs at high redshift are poorly constrained due to the lack of the sample of high-redshift low-mass galaxies caused by insufficient depths of observations.
These observational limits can be resolved by the capability of the Wide Field Camera 3 (WFC3/IR) on board the {\it Hubble} Space Telescope (HST) together with the strong gravitational lensing effect caused by galaxy clusters.
In this manner, the {\it Hubble} Frontier Fields \citep[HFF;][]{Coe+15, Lotz+17} spent more than 800 orbits of the HST to survey six galaxy cluster fields, extending the faint end of the rest-frame ultraviolet (UV) luminosity functions (LFs) to a intrinsic UV magnitude $M_\UV\approx-14~\mathrm{mag}$ at $z\gtrsim6$ \citep{Finkelstein+15, Laporte+16, McLeod+16, Livermore+17, Ishigaki+18, Atek+18, Bhatawdekar+18}.
Here, one should study the GSMFs complementary to the UV LFs that represent the fundamental physical quantity of the stellar mass.

The lensing effect has brought further benefit to studies of galaxy structures and morphologies.
Lensed galaxies are stretched along critical curves, allowing the structures to be studied at high spatial resolution.
Note that this can be fully exploited with the high resolving power of the WFC3/IR.
Owing to magnification by the lensing effects, understanding the size evolution has been significantly advanced by recent studies \citep[e.g.,][]{Shibuya+15, Bouwens+17, Kawamata+18}.
\citet{Bouwens+17} compare the effective radii $R_\mathrm{e}$ and the stellar masses $M_\star$ of the $z\sim6-8$ galaxies to study the analogy of the high-redshift galaxies to the local stellar systems, reporting that some galaxies have the values of $R_\mathrm{e}$ and $M_\star$ comparable to those of the globular clusters (GCs), super star complexes, or star cluster complexes in the local universe.
Here, \citet{Bouwens+17} assume $M_\star$-$M_\UV$ relations, i.e. stellar mass to UV luminosity $L_\UV$ ratios $M_\star/L_\UV$, to estimate $M_\star$ with $M_\UV$ measurements.
In \citet{Bouwens+17}, the assumed $M_\star$-$M_\UV$ relations are based on two representative models with star-formation duration time of 10 and 100 Myr (including the time difference by a factor of $\sim 4$) that have not been tested with observational data yet.
The $M_\star$-$M_\UV$ relation should be determined by observations with no such assumptions, and the $R_\mathrm{e}$-$M_\star$ relation should be studied.

The structure of this paper is as follows.
We describe the data and sample in detail in Section \ref{sec:data}.
The $M_\star$-$M_\UV$ relations are obtained by the staking analysis and the spectral energy distribution (SED) fitting technique in Section \ref{sec:sedfitting}.
In Section \ref{sec:results}, we show the results and the discussions on the GSMFs and the $R_\mathrm{e}$-$M_\star$ relations of the galaxies.
Finally, we give a summary in Section \ref{sec:summary}.

Throughout this paper, we adopt a cosmology with $\Omega_{\mathrm{m},0}=0.3$, $\Omega_{\Lambda,0}=0.7$, and $H_0=70~\mathrm{km}~\mathrm{s}^{-1}~\mathrm{Mpc}^{-3}$.
Magnitudes are given in the AB system \citep{Oke+83}.
Densities and sizes of the galaxies are measured in comoving and physical scales, respectively.
We adopt the \citet{Chabrier03} initial mass function (IMF) in a mass range of $0.1-100~\MS$ to estimate stellar masses.
In this paper, all of the stellar masses taken from the previous studies are converted to those estimated with the \citet{Chabrier03} IMF.

\section{Data and Samples} \label{sec:data}

\subsection{HST Data and Samples} \label{subsec:hst}

In our analysis, we make use of the image mosaics obtained in the HFF program, which targets six cluster fields---Abell 2744, MACS J0416.1$-$2403, MACS J0717.5$+$3745, MACS J1149.6$+$2223, Abell S1063, and Abell 370 (hereafter A2744, M0416, M0717, M1149, A1063, and A370, respectively)---and their accompanying six parallel fields.
All of the 12 fields were observed with the three bands of the Advanced Camera for Surveys (ACS) and four bands of the WFC3/IR; F435W (\bnB), F606W (\bnV), F814W (\bni), F105W (\bnY), F125W (\bnJ), F140W (\bnJH), and F160W (\bnH).
We utilize the drizzled and weight images that were produced by \citet{Shipley+18} in the manner summarized below.
\footnote{\url{http://cosmos.phy.tufts.edu/~danilo/HFF/Download.html}}
First, the HFF v1.0 images were downloaded from the MAST archive.
\footnote{\url{http://www.stsci.edu/hst/campaigns/frontier-fields/}}
The point-spread functions (PSFs) of these images were homogenized to those of the {\bnH} images.
The PSF FWHM of the homogenized images is $\approx0\farcs18$.
Second, the bright cluster galaxies (bCGs) were modeled and subtracted from the images to avoid the diffuse intracluster light (ICL) in photometry of the background faint sources.
All of the images have a pixel scale of $0\farcs06$.
We correct for the Galactic extinction using the values given by \citet{Shipley+18}.

We divide each image into $3\times3$ grid cells, and measure the limiting magnitude in each cell ($\sim1~\mathrm{arcmin}^2$). This is because limiting magnitudes are not homogeneous due to the ICL \citep[e.g.,][]{Montes+14, Ishigaki+15, Kawamata+16}.
The $5\sigma$ limiting magnitudes in the {\bnH} band images are $\approx28.4-29.2~\mathrm{mag}$ in a $0\farcs35$-diameter circular aperture.

We use the galaxy sample selected by \citet{Kawamata+18}, which consists of 350 $i$-, 64 $Y$-, and 39 $Y\!J$-dropouts (or $z\sim6-7,~z\sim8,~z\sim9$ Lyman break galaxies, respectively).
Photometry is reperformed in the following way.
First, we measure the aperture magnitude $m_\aper$ with a diameter of $D_\aper\equiv0\farcs35$ at the position of the dropouts, using the IRAF task {\tt phot} \citep[][\citeyear{Tody93}]{Tody86}.
Second, we apply an aperture correction.
To evaluate the aperture correction term, we create a median-stacked {\bnJ}-band image of the $i$-dropouts with PSF homogenization, and measure the aperture flux of the stacked dropout with changing the aperture diameter.
We define $D_{99}$ by the diameter of the aperture that includes $\geq99~\%$ of the total flux is included.
The aperture correction term, $c_\aper$, is defined as $c_\aper\equiv m_\aper(D_\aper)-m_\aper(D_{99})$, where $m_\aper(d)$ is a aperture magnitude for a given diameter $d$.
For {\bnB} and {\bnV} bands, $c_\aper$ is defined as the same value defined in the {\bni} band, because $c_\aper$ cannot be estimated due to low signal-to-noise (S/N) ratios in the {\bnB} and {\bnV} bands.
The values of $c_\aper$ are $0.98,~0.98,~0.98,~0.93,~0.95,~0.94$, and $0.90~\mathrm{magnitudes}$ in the {\bnB}, {\bnV}, {\bni}, {\bnY}, {\bnJ}, {\bnJH}, and {\bnH} bands, respectively.
Finally, we estimate the total magnitude $m_\tot$ as $m_\tot=m_\aper(D_\aper)-c_\aper$.

In our photometry, some dropouts may be detected at the bands blueward of the Lyman break.
We remove the dropouts with a $\geq2\sigma$ detection either in {\bnB} or {\bnV} band for $i$-dropouts, and in {\bnB}, {\bnV}, or {\bni} band for $Y$- and $Y\!J$-dropouts. For our dropouts, we apply criteria of $\geq3\sigma$ detections in {\bnJ}, {\bnJH}, and {\bnH} bands for $i$-, $Y$-, and $Y\!J$-dropouts, respectively.
Our final sample consists of 267 $i$-dropouts, 54 $Y$-dropouts, and 36 $Y\!J$-dropouts (357 in total).

\subsection{VLT and Keck Data} \label{subsec:vlt}

We use the {\bnKs} band images obtained by the K-band Imaging of the Frontier Fields (KIFF) program \citep{Brammer+16}.
The deep {\bnKs} images of the VLT/HAWK-I (Keck-I/MOSFIRE) are available in both the cluster and the parallel fields of A2744, M0416, A1063, and A370 (M0717 and M1149).
The HAWK-I and MOSFIRE images have the PSF FWHM of $\approx0\farcs4-0\farcs5$.
We utilize the drizzled and weight images of \citet{Shipley+18}, whose details are described in Section \ref{subsec:hst}.
The pixel scale of the images is $0\farcs06$ that matches to that of the HST images.
Photometry and limiting-magnitude measurements are conducted in the same manner as those in Section \ref{subsec:hst}, but with $D_\aper=0\farcs6$ and $c_\aper=0.98$.
The $5\sigma$ limiting magnitudes are $\approx25.3-26.2~\mathrm{mag}$.

\subsection{{\it Spitzer} Data} \label{subsec:irac}

We utilize the drizzled and weight images of ch1 ($3.6~\micro$) and ch2 ($4.5~\micro$) of the IRAC.
The photometric data that were taken by 2016 December are combined by \citet{Shipley+18}, who also applies bCG subtraction and Galactic extinction correction.
In both ch1 and ch2,  the PSF FWHM is $\approx1\farcs7-2\farcs0$, and pixel scales are $0\farcs3$.
Photometry and limiting-magnitude measurements are conducted in the same manner as described in Section \ref{subsec:hst}, but with $D_\aper=3\farcs0$ for both ch1 and ch2, and $c_\aper=0.52,~0.55$ for ch1 and ch2, respectively \citep{Ono+10a}.
The $5\sigma$ limiting magnitudes are $\approx24.5-25.5~\mathrm{mag}$.

\subsection{Lens Models} \label{subsec:lens}

We apply the best-fit magnification factors $\mu$ to estimate the intrinsic magnitudes and stellar masses of the dropouts.
The magnification factors that we use are calculated with the {\tt glafic} \citep{Oguri10} parametric models by \citet{Kawamata+16, Kawamata+18}.
\footnote{\url{https://archive.stsci.edu/prepds/frontier/lensmodels/}}
\citet{Priewe+17} estimate the uncertainties of the eight mass models of M0416, using a fractional normalized median absolute deviation, which is defined as
\begin{equation}
 \text{fNMAD}\equiv1.4826\times\mathrm{median}|\mu_\mathrm{m}(\bm{\theta})-\tilde{\mu}|/\tilde{\mu}.
\end{equation}
Here $\mu_\mathrm{m}(\bm{\theta})$ is the magnification factor calculated with a given model m at a given position $\bm{\theta}$, and $\tilde{\mu}$ is the median of the magnification factors.
It is found that $\text{fNMAD}\approx0.3~(0.7)$ at $\tilde{\mu}\approx2~(40)$ in M0416, which implies that magnification factors of different models differ only by a factor of $\approx2$.
The magnification uncertainties do not change our conclusions (see also \citealt{Ishigaki+15,Meneghetti+17,Kawamata+18}).

\section{Stellar Population of the High-$z$ Galaxies} \label{sec:sedfitting}

\subsection{Stacking Analysis} \label{subsec:stacking}

To create images with high S/N ratios, we conduct a stacking analysis in the following way.
We take {\bnJ}, {\bnJH}, and {\bnH} band magnitudes as the rest-frame UV apparent magnitudes $m_\UV$ of the $i$-, $Y$-, and $Y\!J$-dropouts, respectively.
The intrinsic absolute magnitude of a dropout is calculated via
\begin{equation}
 M_\UV=m_\UV-5\log\left[\dfrac{d_\mathrm{L}(z)}{10~\mathrm{pc}}\right]+2.5\log[\mu(z)]+2.5\log(1+z),
\end{equation}
where $d_\mathrm{L}(z)$ is the luminosity distance to the dropout, and $\mu(z)$ is the best-fit magnification factor at the position of the dropout (see Section \ref{subsec:lens}).
We apply $z=6$, $8$, and $9$ for the $i$-, $Y$-, and $Y\!J$-dropouts, respectively, for consistency with the assumption in the size measurements \citep{Kawamata+18}.
We divide our dropout samples into subsamples by the values of $M_\UV$ at $z\sim6-7$, $8$, and $9$.
A summary of the subsamples is shown in Table \ref{tab:subsamples}.

In each band, we cut out images centered at the positions of dropouts, and divide the pixel counts by $\mu$ of the dropouts.
Here we assume that the values of $\mu$ do not change within sizes of the dropout galaxies, because the sizes of the dropout galaxies are too small to significantly impact on the dropout galaxy fluxes.
We then median-stack the images of the dropouts for each subsample with {\tt iraf} task {\tt imcombine}.

\begin{figure}
\epsscale{1.2}
\plotone{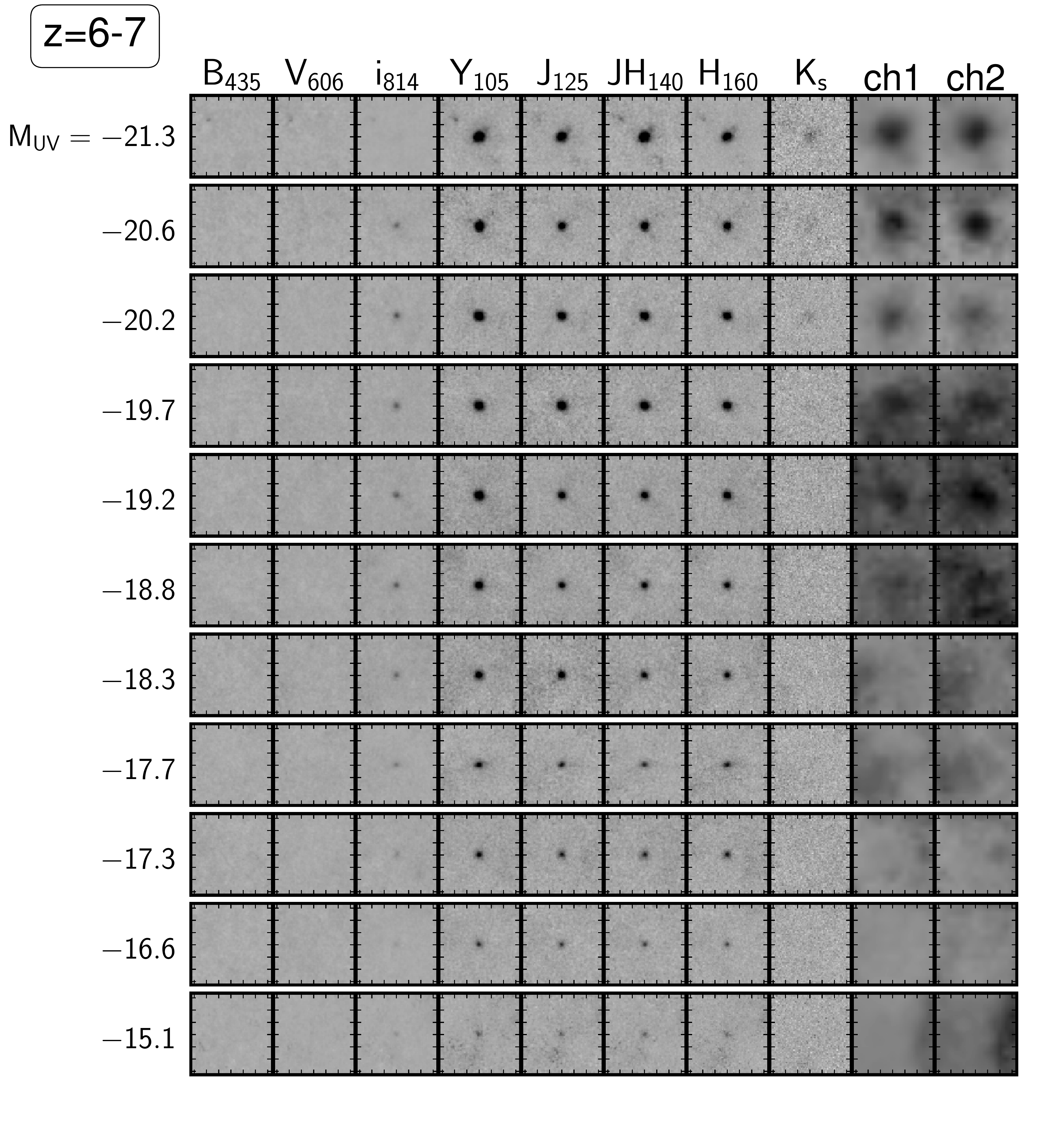}
\caption{Stacked images of the $z\sim6-7$ subsamples for each band. The image size is $4''\times4''$.}
\label{fig:stackimage_z6-7}
\end{figure}

\begin{figure}
\epsscale{1.2}
\plotone{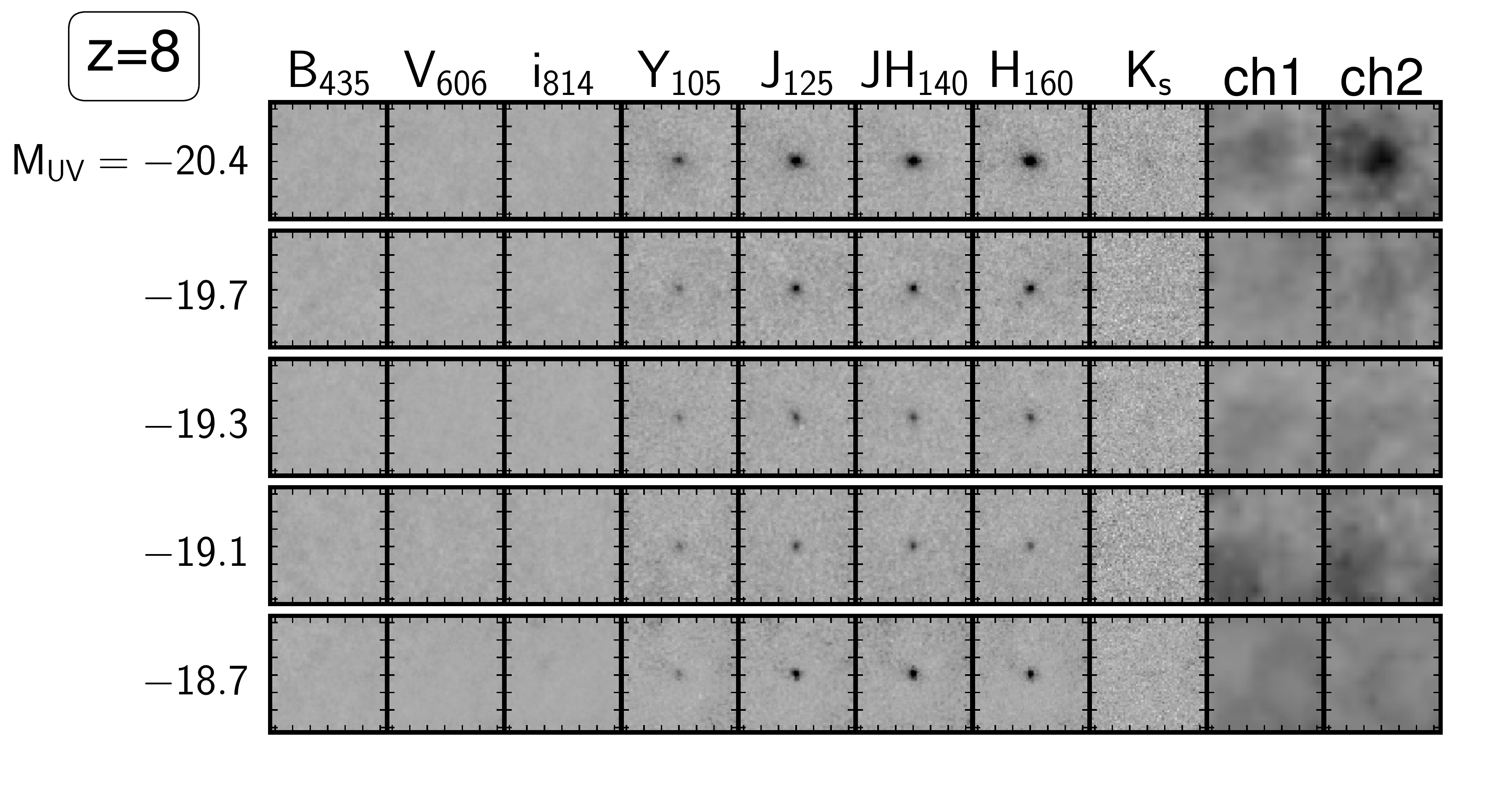}
\caption{Same as Figure \ref{fig:stackimage_z6-7}, but for the $z\sim8$ subsamples.}
\label{fig:stackimage_z8}
\end{figure}

\begin{figure}
\epsscale{1.2}
\plotone{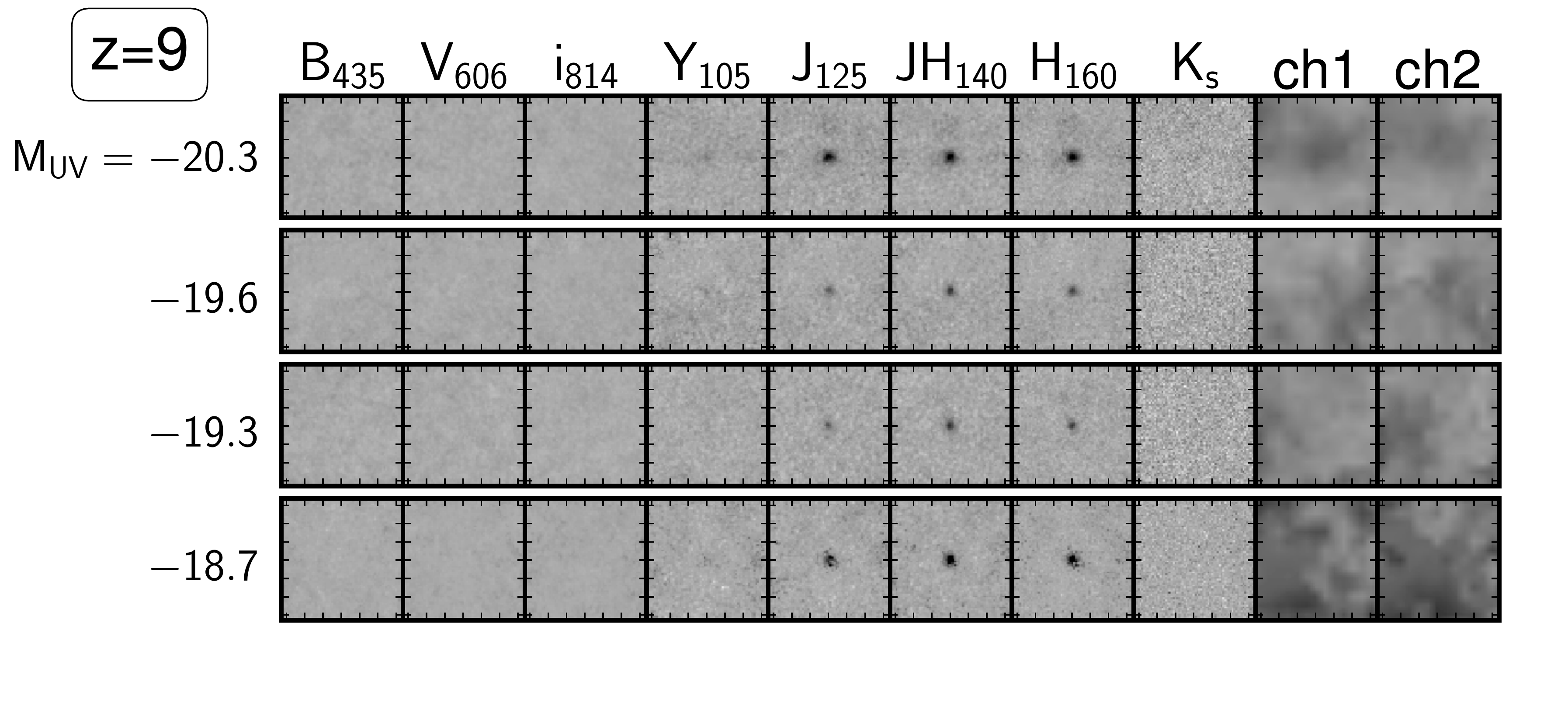}
\caption{Same as Figure \ref{fig:stackimage_z6-7}, but for the $z\sim9$ subsamples.}
\label{fig:stackimage_z9}
\end{figure}

Figures \ref{fig:stackimage_z6-7}, \ref{fig:stackimage_z8}, and \ref{fig:stackimage_z9} show the stacked images of the $z\sim6-7$, $8$, and $9$ subsamples, respectively.
Magnitudes for the subsamples are measured in the same manner as those for the individual dropouts in Section \ref{subsec:hst}.
To estimate uncertainties of the total fluxes (or the total magnitudes), we first follow the three steps; 1) randomly selecting positions in the sky area of the grid cell where the subsamples are located, 2) generating median-stacked sky noise images, and 3) performing aperture photometry on the sky noise images with the aperture correction to obtain the total flux $f_i$.
The steps 1)-3) are repeated for 100 times.
We make a histogram of $f_i$, and fit the histogram with a Gaussian profile.
We regard the standard deviation divided by the median magnification factor of the subsample as the uncertainty of the total flux.

\begin{deluxetable}{rcc}
\tablecaption{Summary of the subsamples.
\label{tab:subsamples}}
\tablehead{
\colhead{Subsample} & \colhead{Threshold} & \colhead{$N$}
}
\colnumbers
\startdata
$z\sim6-7,~M_\UV=-21.3$	&	$\hspace{12.7mm}M_\UV\leq-21.0$	&	$4$		\\
$-20.6$					&	$-21.0\leq M_\UV\leq-20.5$		&	$7$		\\
$-20.2$					&	$-20.5\leq M_\UV\leq-20.0$		&	$19$		\\
$-19.7$					&	$-20.0\leq M_\UV\leq-19.5$		&	$43$		\\
$-19.2$					&	$-19.5\leq M_\UV\leq-19.0$		&	$58$		\\
$-18.8$					&	$-19.0\leq M_\UV\leq-18.5$		&	$64$		\\
$-18.3$					&	$-18.5\leq M_\UV\leq-18.0$		&	$30$		\\
$-17.7$					&	$-18.0\leq M_\UV\leq-17.5$		&	$18$		\\
$-17.3$					&	$-17.5\leq M_\UV\leq-17.0$		&	$15$		\\
$-16.6$					&	$-17.0\leq M_\UV\leq-16.0$		&	$10$		\\
$-15.1$					&	$-16.0\leq M_\UV\hspace{12.7mm}$	&	$5$		\\ \tableline
$z\sim8,~M_\UV=-20.4$		&	$\hspace{12.7mm}M_\UV\leq-20.0$	&	$13$		\\
$-19.7$					&	$-20.0\leq M_\UV\leq-19.5$		&	$10$		\\
$-19.3$					&	$-19.5\leq M_\UV\leq-19.25$		&	$10$		\\
$-19.1$					&	$-19.25\leq M_\UV\leq-19.0$		&	$11$		\\
$-18.7$					&	$-19.0\leq M_\UV\hspace{12.7mm}$	&	$10$		\\ \tableline
$z\sim9,~M_\UV=-20.3$		&	$\hspace{12.7mm}M_\UV\leq-19.9$	&	$8$		\\
$-19.6$					&	$-19.9\leq M_\UV\leq-19.5$		&	$9$		\\
$-19.3$					&	$-19.5\leq M_\UV\leq-19.1$		&	$9$		\\
$-18.7$					&	$-19.1\leq M_\UV\hspace{12.7mm}$	&	$10$
\enddata
\tablecomments{Columns: (1) subsample name that indicates the redshift and the median value of $M_\UV$; (2) threshold of the subsample; (3) number of the dropouts in the subsample.}
\end{deluxetable}

\subsection{SED Modeling} \label{subsec:sedfitting}

To investigate the typical stellar mass $M_\star$ for a given $M_\UV$, we conduct the SED fitting method.
We use the {\tt BEAGLE} tool \citep{Chevallard+16}, which is based on a recent version of the stellar population models of \citet{Bruzual+03} and the photoionization models of \citet{Gutkin+16} that are computed with {\tt CLOUDY} \citep{Ferland+13}.
The intergalactic medium (IGM) absorptions follow the models of \citet{Inoue+14}.
We adopt the Calzetti et al. (1994) law to the models for dust attenuation.

There are six free parameters in the fitting: i) redshift $z$ of the galaxy, ii) galaxy age $t_\text{age}$, iii) galaxy-wide ionization parameter $U_\text{ion}$, iv) total mass of the formed stars $M_\star$, v) stellar metallicity $Z$, and vi) {\bnV}-band dust attenuation optical depth $\tau_\mathrm{V}$.
We assume uniform prior probability distribution functions (PDFs) in the range of
\begin{align}
z_\phot^\median-0.1 & \leq z\leq~z_\phot^\median+0.1, \\
1~\mathrm{Myr} & \leq t_\text{age}\leq t_\text{univ}(z=z_\phot^\median)-t_\text{univ}(z=15), \\
-3 & \leq\log{U_\text{ion}}\leq-1, \\
6 & \leq\log(M_\star/\MS)\leq11, \\
-2 & \leq(\log{Z}/\ZS)\leq0.2, \quad \text{and} \\
0 & \leq\log{\tau_\mathrm{V}}\leq2
\end{align}
for the parameters i), ii), iii), iv), v), and vi), respectively.
Here, $z_\phot^\median$ is the median of the photometric redshifts $z_\phot$ of the subsample whose values are taken from \citet{Kawamata+18}.
We omit $z_\phot$ values that are lower than $4$ to derive $z_\phot^\median$ in the same manner as the estimations of the median photometric redshifts \citet{Kawamata+18}.
The age of the universe at $z$ is represented by $t_\text{univ}(z)$.
We assume a constant star formation history.
The interstellar medium metallicity is assumed to equal the stellar metallicity.
The dust-to-metal ratio is fixed to 0.3 \citep[e.g.,][]{De-vis+17}.

\begin{figure}
\epsscale{1.2}
\plotone{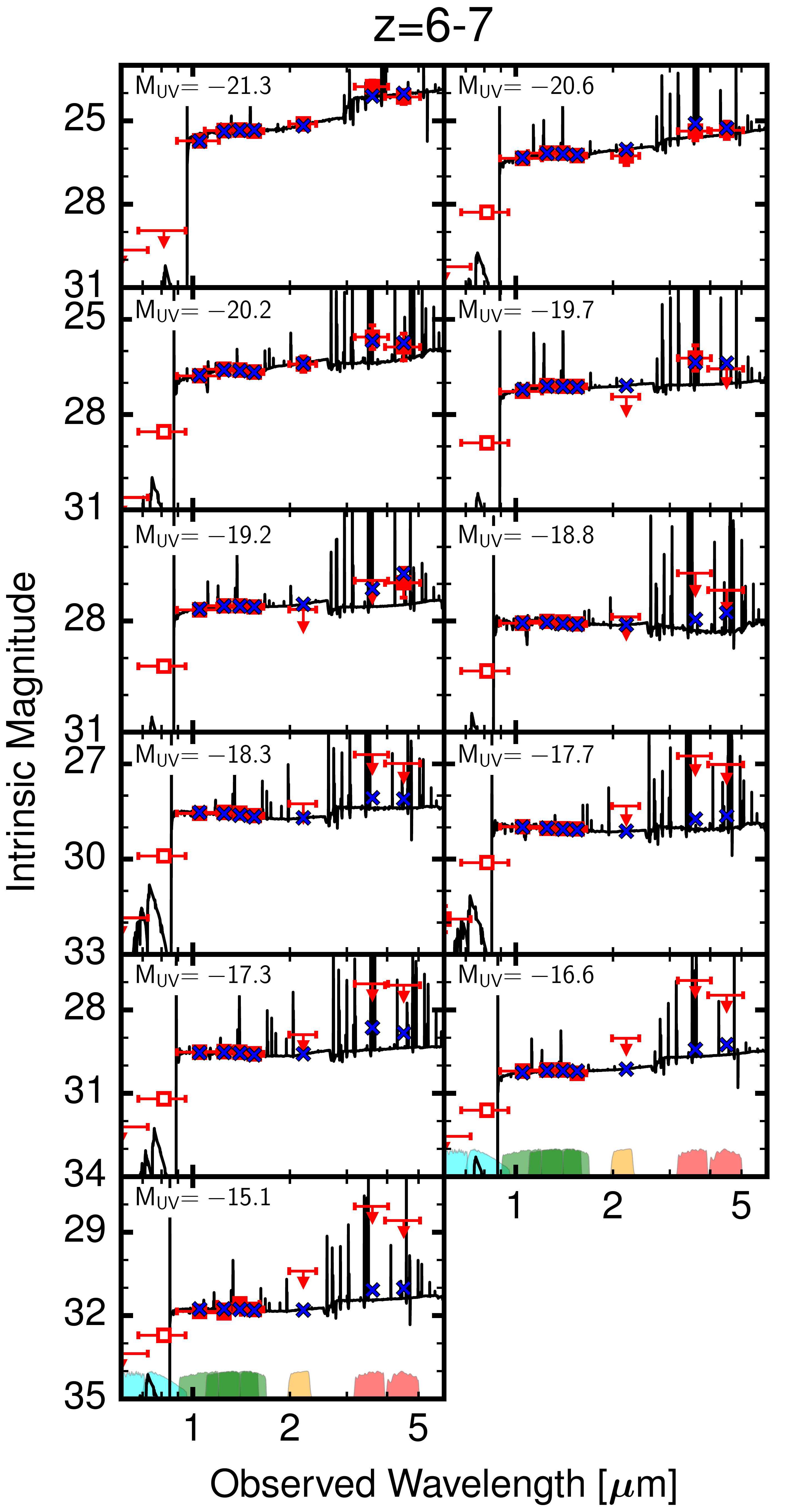}
\caption{Intrinsic SEDs of the $z\sim6-7$ subsamples. The subsample names are denoted at the upper-left corner in each panel. The red filled squares show the photometric data obtained in Section \ref{subsec:stacking}. The red open squares at {\bni} band are the same as the red filled squares, but for the photometry that are not included in the likelihood calculation (see text). The horizontal and vertical error bars represent the wavelength range of the filters and the $1\sigma$ uncertainties, respectively. The down arrows indicate $2\sigma$ upper limits for bands with $\SN<2$. The black lines represent the best-fit SEDs, while the blue crosses show the bandpass-averaged magnitudes predicted from the best-fit SEDs. In the bottom panels, the color shades denote the normalized filter throughputs of {\bnB}, {\bnV}, {\bni} (cyan), {\bnY}, {\bnJ}, {\bnJH}, {\bnH} (green), {\bnKs} (orange), ch1, and ch2 (red).}
\label{fig:fullsed_z6-7}
\end{figure}

\begin{figure}
\epsscale{1.2}
\plotone{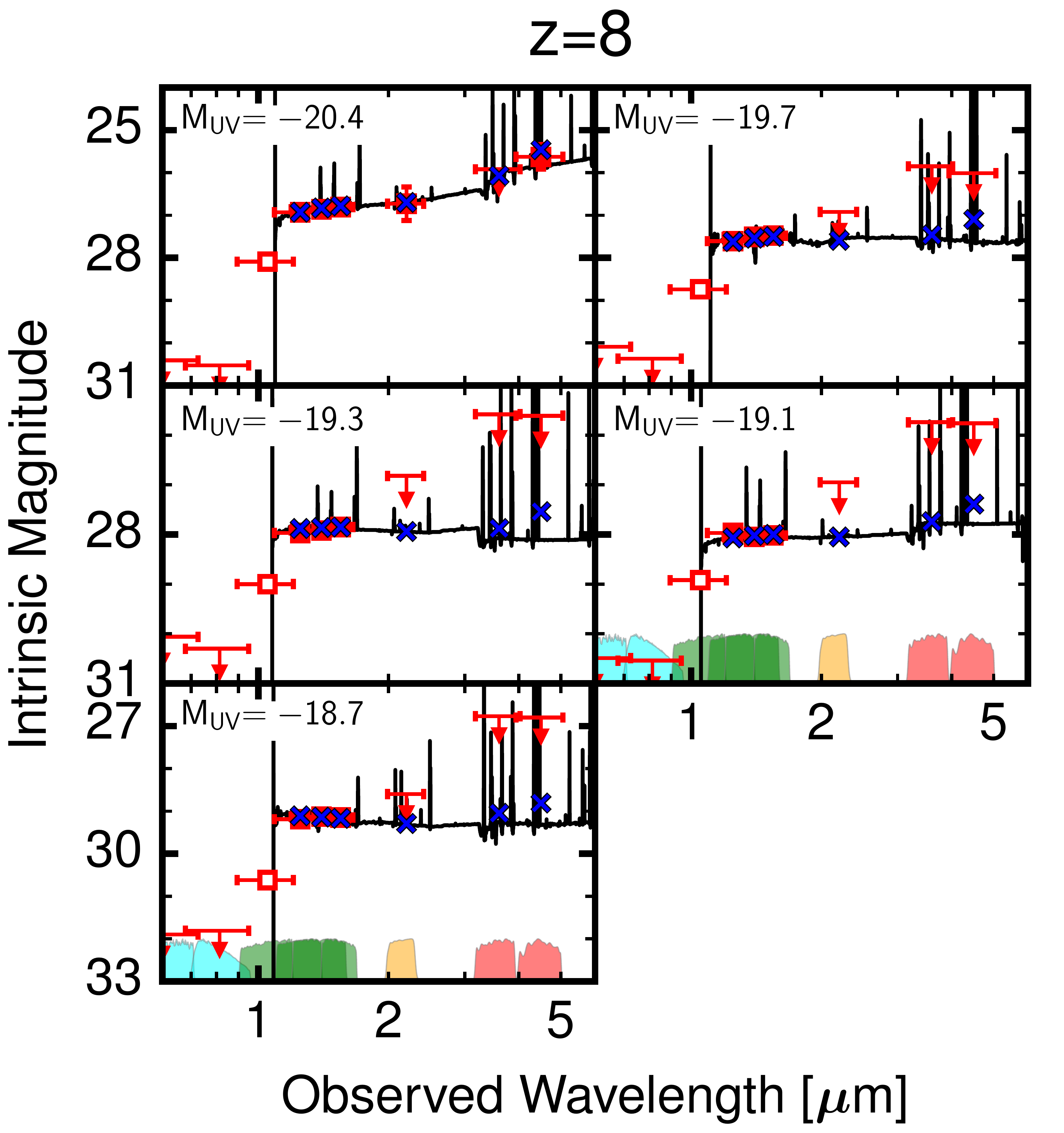}
\caption{Same as Figure \ref{fig:fullsed_z6-7}, but for the $z\sim8$ subsamples.}
\label{fig:fullsed_z8}
\end{figure}

\begin{figure}
\epsscale{1.2}
\plotone{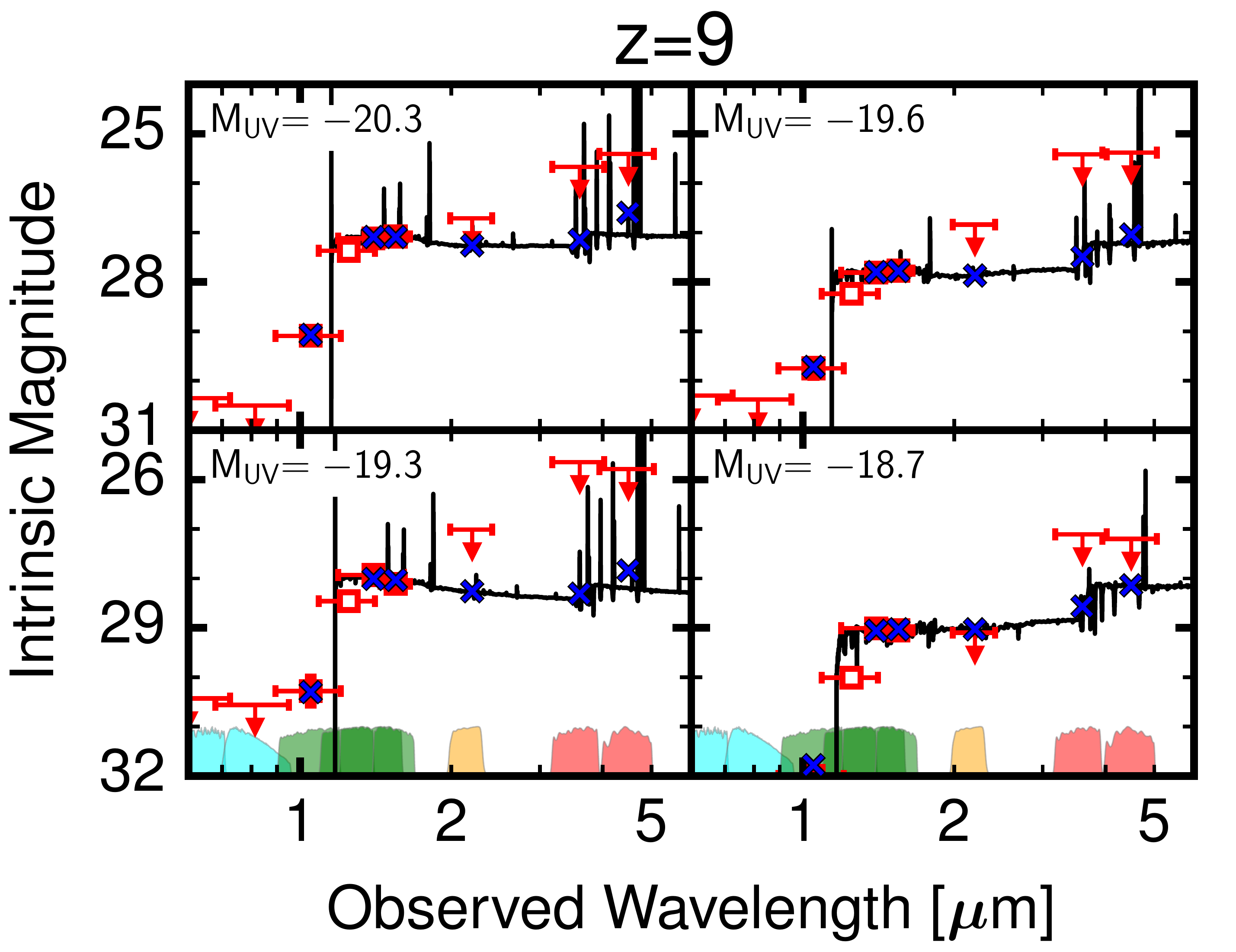}
\caption{Same as Figure \ref{fig:fullsed_z6-7}, but for the $z\sim9$ subsamples.}
\label{fig:fullsed_z9}
\end{figure}

\begin{figure}
\epsscale{1.2}
\plotone{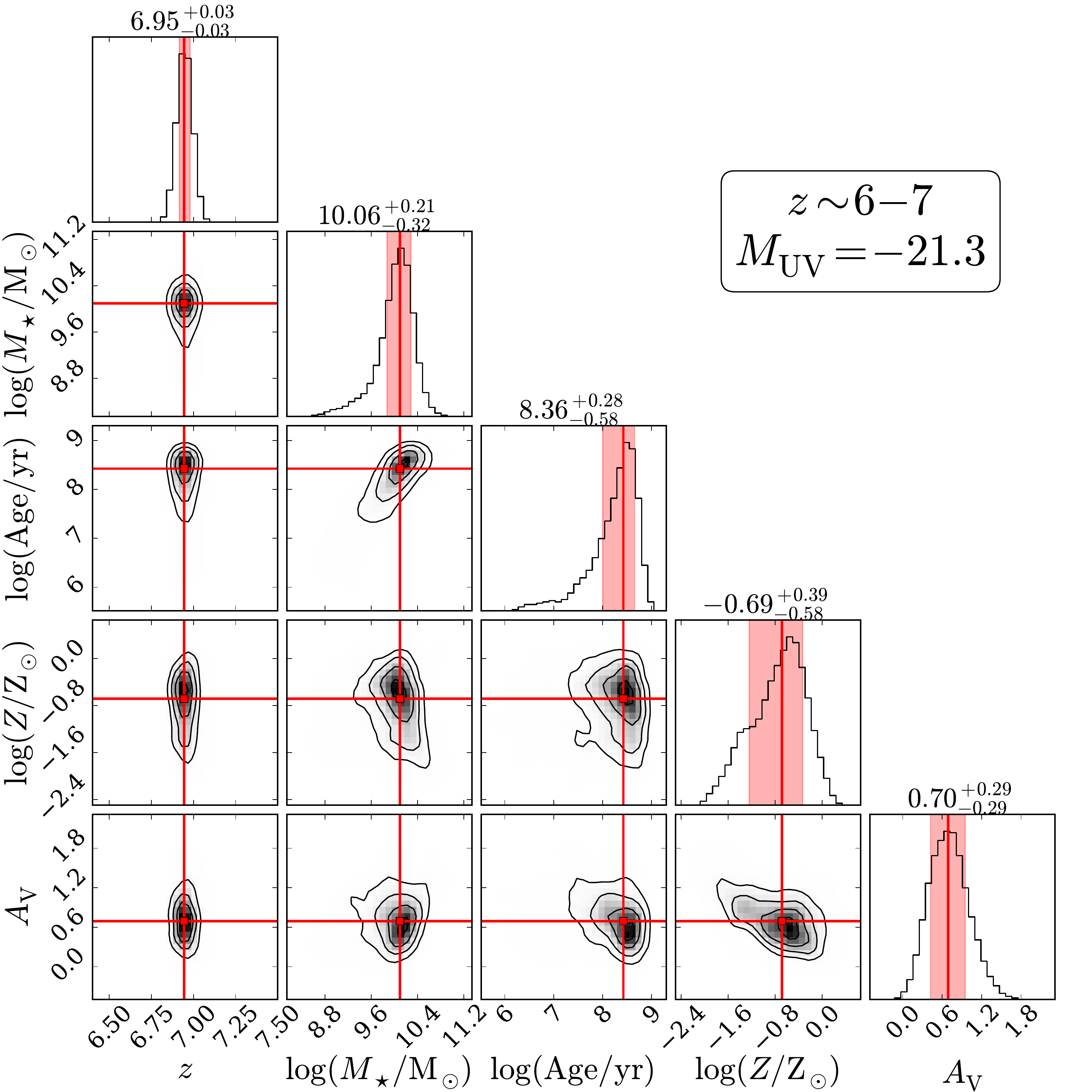}
\caption{Posterior PDFs of the parameters for the $z\sim6-7,~M_\UV=-21.3$ subsample. The diagonal panels show the marginal distributions of $z$, $\log(M_\star/\MS)$, $\log(t_\text{age})$, $\log(Z/\ZS)$, and $A_\mathrm{V}$. Note that $A_\mathrm{V}$ is derived from $\tau_\mathrm{V}$. The red vertical lines and the shades represent the median values and the $1\sigma$ intervals, respectively. The off-diagonal panels represent the joint distributions for the sets of two parameters. The inner, middle, and outer contours correspond to $1\sigma$, $2\sigma$, and $3\sigma$ intervals, respectively.}
\label{fig:tri_eg}
\end{figure}

The posterior PDF $P(\bm{\Theta}|\bm{D},\mathsf{H})$ of a given parameter set $\bm{\Theta}=\{z,t_\text{age},U_\text{ion},M_\star,Z,\tau_\mathrm{V}\}$ of a model $\mathsf{H}$ is calculated based on the Bayes' theorem
\begin{equation}
P(\bm{\Theta}|\bm{D},\mathsf{H})=\dfrac{P(\bm{\Theta}|\mathsf{H})P(\bm{D}|\bm{\Theta},\mathsf{H})}{\int\mathrm{d}\bm{\Theta}P(\bm{\Theta}|\mathsf{H})P(\bm{D}|\bm{\Theta},\mathsf{H})}
\end{equation}
\citep[e.g.,][]{Jeffreys61}, where $\bm{D}$ is the data set, i.e., the fluxes in the {\bnB}, {\bnV}, {\bni}, {\bnY}, {\bnJ}, {\bnJH}, {\bnH}, {\bnKs}, ch1, and ch2 bands.
The likelihood function of $\bm{\Theta}$, $\mathcal{L}(\bm{\Theta})\equiv P(\bm{D}|\bm{\Theta},\mathsf{H})$, is defined via
\begin{equation}
\ln\mathcal{L}(\bm{\Theta})=-\dfrac{1}{2}\sum_k\left[\dfrac{f_k-\hat{f}_k(\bm{\Theta})}{\sigma_k}\right]^2.
\end{equation}
Here $f_k$, $\hat{f}_k(\bm{\Theta})$, and $\sigma_k$ are the observed flux, the flux predicted by the parameter set $\bm{\Theta}$, and the flux uncertainty, respectively.
The subscript $k$ runs over all of the bands but the {\bni}, {\bnY}, and {\bnJ} band for $z\sim6-7$, $8$, and $9$ subsamples, respectively.
We do not include the {\bni}, {\bnY}, and {\bnJ} band for $z\sim6-7$, $8$, and $9$ subsamples, respectively.
This is because we need to avoid the broadband photometry contaminated by unknown Ly$\alpha$ emission and IGM absorption effects.
The value of $\sigma_k$ is defined as
\begin{equation}
\sigma_k\equiv\sqrt{(\sigma_k^\text{obs})^2+(\sigma_0f_k)^2}, \label{eqn:sigma}
\end{equation}
where $\sigma_k^\text{obs}$ is the observational uncertainty estimated in Section \ref{subsec:stacking}, and $\sigma_0$ is the relative systematical uncertainty, e.g., errors in background subtraction, flux calibration, and model predictions \citep{Brammer+08,Dahlen+13,Acquaviva+15,Chevallard+16}.
We define $\sigma_0=0.04~(0.05)$ for the HST and {\bnKs} (IRAC) bands, applying the values recommended in the user manual of $\tt BEAGLE$ \citep{Chevallard+16}.
If the S/N ratio of the stacked image in the band $k$ is less than $2$, we substitute $f_k$ values with $0$.
The posterior PDFs are efficiently sampled by the Nested Sampling algorithm implemented in the {\tt MULTINEST} tool \citep{Feroz+08, Feroz+09}.
We present the best-fit SEDs with the data photometries for $z\sim6-7$, $8$, and $9$ subsamples in Figure \ref{fig:fullsed_z6-7}, \ref{fig:fullsed_z8}, and \ref{fig:fullsed_z9}, respectively.
An example of the posterior PDFs of the parameters is shown in Figure \ref{fig:tri_eg}.

\subsection{Stellar Mass to UV Luminosity Relations} \label{subsec:masslumi}

\begin{figure*}
\epsscale{1.2}
\plotone{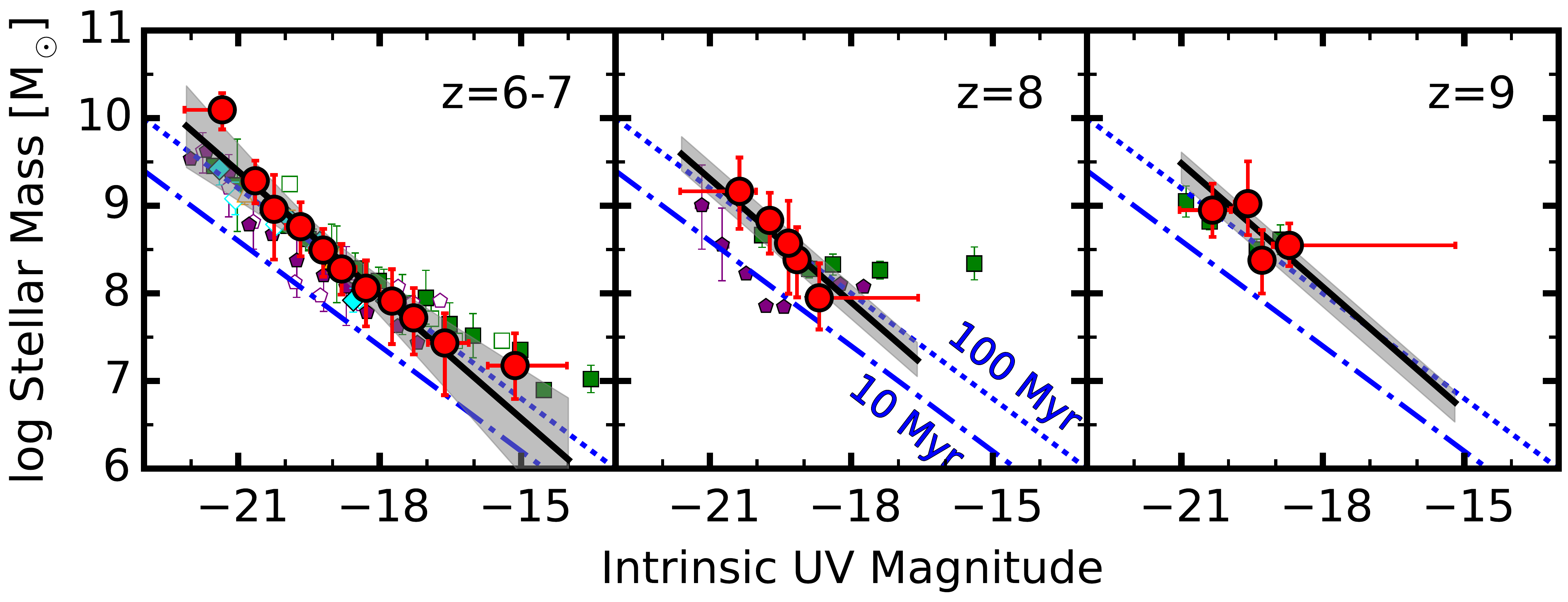}
\caption{Distribution of $M_\star$ versus $M_\UV$ of the $z\sim6-7$ (left), $8$ (middle), and $9$ (right) subsamples. The red filled circles represent our results.  The black lines and the gray shades denote the best-fit relations and the $1\sigma$ uncertainties, respectively. The blue dashed (dotted) lines show the $M_\star$-$M_\star$ relation of the representative model of the 10 (100) Myr star-formation duration time that is assumed in \citet{Bouwens+17}. The other data points are taken from the previous studies, \citet[][orange triangles]{Stark+13}, \citet[][cyan diamonds]{Duncan+14}, \citet[][purple pentagons]{Song+16}, and \citet[][green boxes]{Bhatawdekar+18}. In the left panel, the filled and open symbols show the data at $z\sim6$ and $7$, respectively. All of the IMFs are converted to a \citet{Chabrier03} IMF.}
\label{fig:masslumi}
\end{figure*}

Figure \ref{fig:masslumi} shows the relations between the stellar mass $M_\star$ and the UV magnitude  $M_\UV$.
Here, $M_\star$ and $M_\UV$ represent the median value of the marginal posterior PDF and the median value of the UV magnitudes of the dropouts in the subsample, respectively.
The vertical error bars represent 68 \% confidence intervals, while the horizontal bars show the minimum and maximum values of $M_\UV$ of the dropouts in the subsamples.
Figure \ref{fig:masslumi} indicates that our results are broadly consistent with the previous results.

We fit a linear function
\begin{equation}
 \log(M_\star/\MS)=a_0+a_1(M_\UV+19.5), \label{eqn:mlbest}
\end{equation}
where the intercept $a_0$ and the slope $a_1$ are set as free parameters, to our data in the magnitude range of $-21.0\leq M_\UV\leq-16.0$.
At $z\sim8$ and $9$, we fix $a_1$ to the best-fit value of $a_1$ of $z\sim6-7$, because fitting is unstable due to the lack of data points.
The black solid lines in Figure \ref{fig:masslumi} show the best-fit relations whose parameters are listed in Table \ref{tab:mlbest}.
We find that the best-fit value of $a_1$ at $z\sim6-7$ is $-0.47\pm0.09$ that is similar to those obtained in the previous studies within the $1\sigma$ uncertainties.
This value of $a_1$ is consistent with $-0.4$ that is the slope for the constant $M_\star/L_\UV$ case, indicating that the average $M_\star/L_\UV$ of dropouts at $z\sim6-7$ is nearly constant at $M_\star\sim10^6-10^9~\MS$.
The best-fit values of intercept $a_0$ are $8.67_{-0.11}^{+0.11}$, $8.59_{-0.19}^{+0.18}$, and $8.76_{-0.15}^{+0.13}$ at $z\sim6-7$, $8$, and $9$, respectively, showing no evolution beyond the errors. No evolution of $a_0$ indicates that stellar populations of the dropouts do not significantly change at $z\sim6-9$ in the magnitude range.

In Figure \ref{fig:masslumi}, the blue dashed and dotted lines denote the $M_\star$-$M_\UV$ relations (corresponding to $M_\star/L_{\rm UV}$ ratios) that are assumed by \citet{Bouwens+17} for their $M_\star$ estimations from their $M_\UV$ measurements for $z\sim6-8$ dropouts.
\citet{Bouwens+17} assume two representative model stellar populations whose star-formation duration times of 10 and 100 Myr with no stellar mass/UV luminosity dependence (See Section \ref{sec:intro} for more details).
The assumptions of 10 and 100 Myr correspond to $(a_0,a_1)=(8.0,-0.4)$ and $(8.6,-0.4)$, i.e.
\begin{equation}
 \log(M_\star/\MS)=8.0-0.4(M_\UV+19.5) \label{eqn:ml_B17_10My}
\end{equation}
and
\begin{equation}
 \log(M_\star/\MS)=8.6-0.4(M_\UV+19.5), \label{eqn:ml_B17_100My}
\end{equation}
respectively.
Figure \ref{fig:masslumi} indicates that
our best-fit $M_\star$-$M_\UV$ relations (the black solid lines) are comparable to the model of 100-Myr star-formation duration time (the blue dotted lines). However, our best-fit $M_\star$-$M_\UV$ relations fall above the model of 10-Myr star-formation duration time (the blue dashed lines). By the comparison with our best-fit parameters at $z\sim 6-7$, we find that the parameter set for the blue dashed line ($a_0=8.0$ and $a_1=-0.4$) is ruled out at the $5\sigma$ confidence level.
The parameter set of $(a_0,a_1)=(8.0,-0.4)$ provides the $M_\star$-$M_\UV$ relation $2-7$ times lower than the one of our best-fit parameters at $z\sim 6-9$ in $M_\star$.

\begin{deluxetable}{ccc}
\tablecaption{Best-fit parameters of the $M_\star$-$M_\UV$ relations.
\label{tab:mlbest}}
\tablehead{
\colhead{$z$} & \colhead{$a_0$} & \colhead{$a_1$}
}
\startdata
$6-7$	&	$8.67_{-0.11}^{+0.11}$	&	$-0.47_{-0.09}^{+0.09}$	\\
$8$		&	$8.59_{-0.19}^{+0.18}$	&	$-0.47$ (fixed)			\\
$9$		&	$8.76_{-0.15}^{+0.13}$	&	$-0.47$ (fixed)
\enddata
\end{deluxetable}

\section{Results and Discussion} \label{sec:results}

\subsection{Galaxy Stellar Mass Functions}

\begin{figure*}
\epsscale{1.2}
\plotone{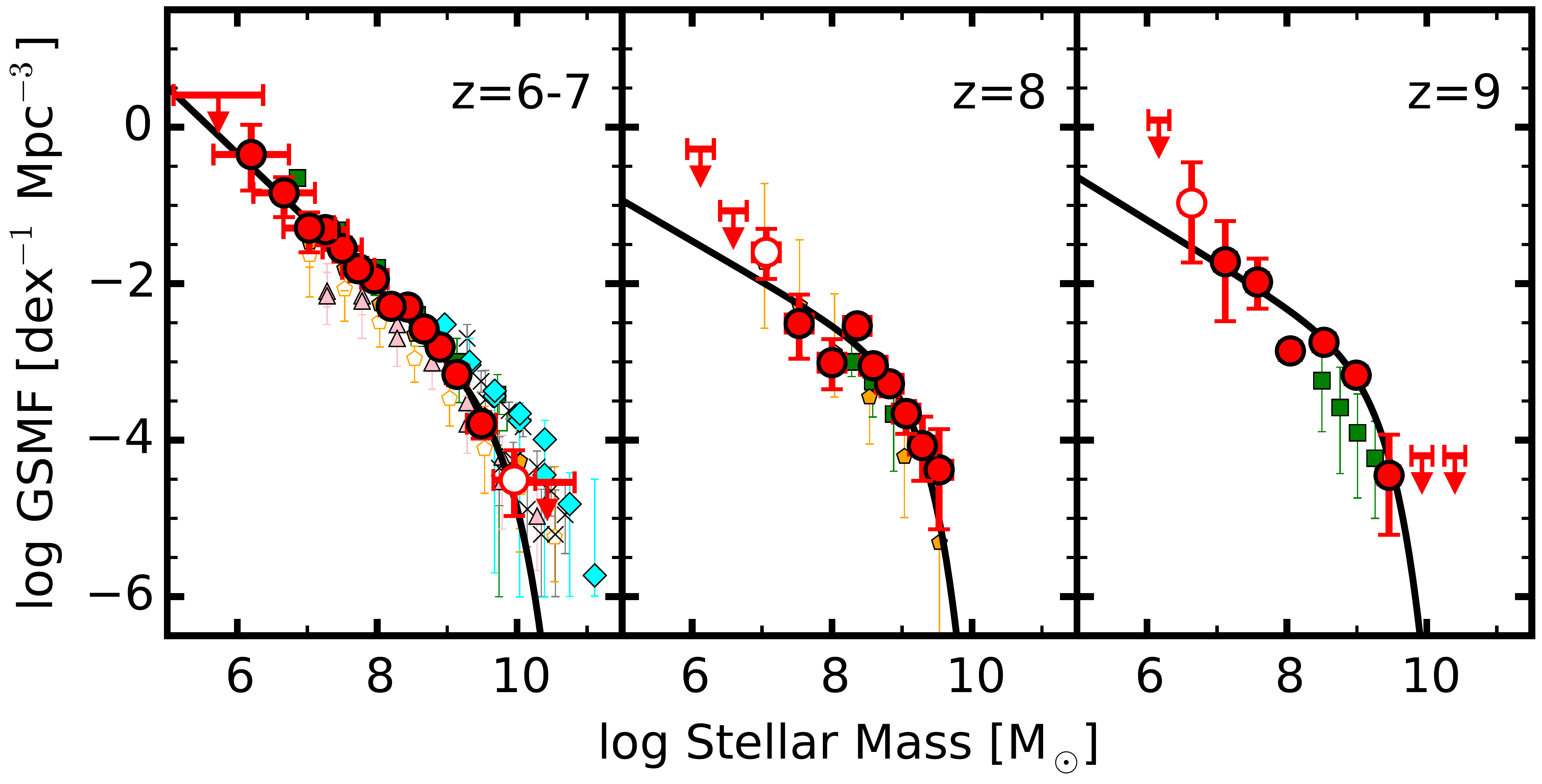}
\caption{GSMFs at $z\sim6-7$ (left), $8$ (middle), and $9$ (right). The red filled circles show our GSMFs. The red open circles are the same as the red filled circles, but for the GSMFs obtained with the extrapolated best-fit $M_\star$-$M_\UV$ relations. We do not include the data of the red open circles for our Schechter function fitting. The down arrows denote the upper limits of the GSMFs. The other data points are taken from the previous studies, \citet[][green boxes]{Bhatawdekar+18}, \citet[][orange pentagons]{Song+16}, \citet[][black crosses]{Grazian+15}, \citet[][cyan diamonds]{Duncan+14}, and \citet[][magenta triangles]{Gonzalez+11}. In the left panel, the filled and open symbols show the GSMFs at $z\sim6$ and $7$, respectively. The black curves represent the best-fit Schechter functions obtained with the measurements of our study and \citet{Song+16}. The stellar masses taken from the previous studies are converted to those estimated with the \citet{Chabrier03} IMF.}
\label{fig:GSMF}
\end{figure*}

To derive the GSMFs, we apply the best-fit $M_\star$-$M_\UV$ relations (Section \ref{subsec:masslumi}) to the LFs of \citet{Ishigaki+18}.
We show our GSMFs in Figure \ref{fig:GSMF} with black dots.
One can find that our GSMFs are broadly consistent with those of previous studies in the range of $M_\star\sim10^7-10^9~\MS$ at $z\sim6-7$.
In the low-mass end, our GSMFs reach $M_\star\sim10^6~\MS$, which is lower than those of the previous studies by $\gtrsim1~\mathrm{dex}$.

We parametrize our GSMFs with the \citet{Schechter76} function,
\begin{equation}
 \Psi(M_\star)=\ln(10)\Psi^*\left(\dfrac{M_\star}{M^*}\right)^{\alpha+1}\exp\left(-\dfrac{M_\star}{M^*}\right). \label{GSMF_Schechter}
\end{equation}
The characteristic stellar mass $M^*$, the low-mass end slope $\alpha$, and the normalization $\Psi^*$ are free parameters whose best-fit values are obtained by the least-squares fitting method.
To improve the statistical accuracy, we fit Schechter functions simultaneously to our GSMFs and the GSMFs derived by \citet{Song+16} who do not use the HFF data but the CANDELS/GOODS \citep{Grogin+11, Koekemoer+11} and the HUDF \citep{Beckwith+06, Illingworth+13} data.
We combine our GSMFs at $z\sim6-7$ and the GSMFs of \citet{Song+16} at $z\sim6$, because the median value of $z_\phot$ of our dropouts is $z=6.3$ that is regarded as $z\sim 6$.
Table \ref{tab:GSMF_Schechpar} summarizes the best-fit Schechter parameters at each redshift.

\begin{deluxetable}{cccc}
\tablecaption{Best-fit parameters of the Schechter function of the GSMFs.
\label{tab:GSMF_Schechpar}}
\tablehead{
\colhead{Ref.}	&	\colhead{$M_\star$}		&	\colhead{$\alpha$}	&	\colhead{$\Psi^*$}	\\
\colhead{}		&	\colhead{$[\log~\MS]$}	&	\colhead{}			&	\colhead{$[10^{-5}~\mathrm{dex}^{-1}~\mathrm{Mpc}^{-3}]$}
}
\startdata
$z\sim6$	&						&						&						\\
K19		&	$9.58_{-0.15}^{+0.23}$	&	$-1.85_{-0.07}^{+0.07}$	&	$18.2_{-9.3}^{+12.7}$	\\
D14		&	$10.87_{-1.06}^{+1.13}$	&	$-2.00_{-0.40}^{+0.57}$	&	$1.4_{-1.4}^{+41.1}$		\\
G15		&	$10.49\pm0.32$		&	$-1.55\pm0.19$			&	$6.19_{-4.57}^{+13.50}$	\\
S16		&	$10.72_{-0.30}^{+0.29}$	&	$-1.91_{-0.09}^{+0.09}$	&	$1.35_{-0.75}^{+1.66}$	\\
B18		&	$10.29_{-0.56}^{+0.48}$	&	$-1.93_{-0.07}^{+0.05}$	&	$5.63_{-3.23}^{+7.12}$	\\ \tableline
$z\sim7$	&						&						&						\\
D14 		&	$10.51$ (fixed)			&	$-1.89_{-0.61}^{+1.39}$	&	$3.6_{-3.5}^{+30.1}$		\\
G15		&	$10.69\pm1.58$		&	$-1.88\pm0.36$			&	$0.57_{-0.57}^{+59.68}$	\\
S16		&	$10.78_{-0.28}^{+0.29}$	&	$-1.95_{-0.18}^{+0.18}$	&	$0.53_{-0.38}^{+1.10}$	\\
B18		&	$10.25_{-0.49}^{+0.45}$	&	$-1.95_{-0.07}^{+0.06}$	&	$2.82_{-1.88}^{+5.16}$	\\ \tableline
$z\sim8$	&						&						&							\\
K19	&	$8.93_{-0.15}^{+0.23}$	&	$-1.52_{-0.26}^{+0.27}$	&	$45.7_{-25.7}^{+33.7}$		\\
S16		&	$10.72_{-0.29}^{+0.29}$	&	$-2.25_{-0.35}^{+0.72}$	&	$0.035_{-0.030}^{+0.246}$	\\
B18		&	$10.48_{-0.75}^{+0.90}$	&	$-2.25_{-0.29}^{+0.23}$	&	$0.089_{-0.074}^{+0.36}$		\\ \tableline
$z\sim9$	&						&						&							\\
K19	&	$9.04_{-0.18}^{+0.47}$	&	$-1.55_{-0.30}^{+0.29}$	&	$60.3_{-46.8}^{+37.5}$		\\
B18		&	$10.45_{-0.85}^{+0.80}$	&	$-2.33_{-0.39}^{+0.30}$	&	$0.06_{-0.06}^{+0.66}$		\\
\enddata
\tablecomments{Reference: K19 $=$ this work, D14 $=$ \citet{Duncan+14}, G15 $=$ \citet{Grazian+15}, S16 $=$ \citet{Song+16}, and B18 $=$ \citet{Bhatawdekar+18}. Here we show the `point source' results from \citet{Bhatawdekar+18}.}
\end{deluxetable}

\subsection{Galaxy Stellar Mass Density}

We derive the galaxy stellar mass densities (GSMDs) at $z\sim6-9$.
We integrate our best-fit Schechter functions of the GSMFs over the stellar mass range of $10^8\leq M_\star/\MS\leq10^{13}$ that is adopted in previous studies.
Table \ref{tab:GSMD} lists our GSMDs at $z\sim6-9$.
The uncertainties of our GSMDs include the statistical errors and the cosmic variance uncertainties, the latter of which are added in quadrature with the values of $0.3$, $0.4$ and $0.5~\mathrm{dex}$ at $z\sim6-7$, $8$ and $9$, respectively \citep{Robertson+14, Ishigaki+15, Bhatawdekar+18}.

\begin{figure}
\epsscale{1.2}
\plotone{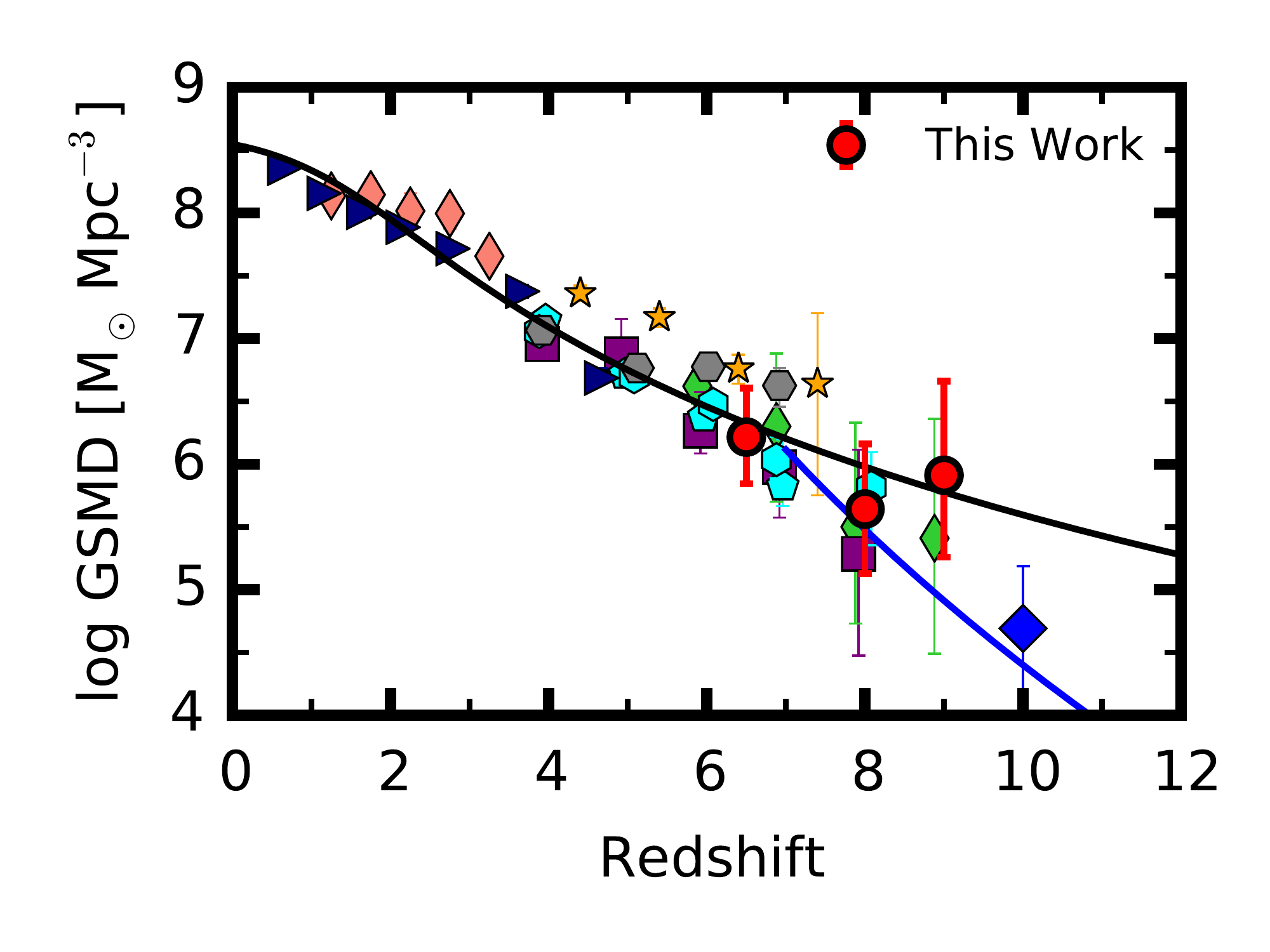}
\caption{GSMDs as a function of redshift. The red circles show our GSMDs. Here the error bars include the statistical errors and cosmic variance uncertainties. Our GSMD at $z\sim6-7$ is plotted at $z=6.5$. The other data points are taken from the previous studies, \citet[][green diamonds]{Bhatawdekar+18}, \citet[][purple boxes]{Song+16}, \citet[][cyan pentagons]{Grazian+15}, \citet[][a blue diamond]{Oesch+14}, \citet[][orange stars]{Duncan+14}, \citet[][cyan hexagons]{Stark+13}, \citet[][gray hexagons]{Gonzalez+11}, \citet[][magenta diamonds]{Mortlock+11}, and \citet[navy triangles]{Elsner+08}. The data points are slightly shifted along the redshift axis if overlapped with other points. The black and blue solid curves indicate the GSMD evolution that is derived as the time integration of the SFRD evolution presented in \citet{Madau+14} and \citet{Oesch+18}, respectively. The stellar masses taken from the previous studies are converted to those estimated with the \citet{Chabrier03} IMF.}
\label{fig:GSMD}
\end{figure}

We show our GSMDs in Figure \ref{fig:GSMD} together with the results obtained in the previous studies.
Our GSMDs are consistent with those of the previous studies within $1 \sigma$ uncertainties.

If the GSMDs and the star-formation rate densities (SFRDs) are accurately measured, the time integral of the SFRDs should be consistent with the GSMDs.
In Figure \ref{fig:GSMD}, we show the evolution of the GSMDs that is derived as the time integration of the SFRDs obtained in \citet{Madau+14} with the black solid curve.
The SFRDs of \citet{Madau+14} declines smoothly by $(1+z)^{-3.9}$ at $z>8$.
Similar trends of the SFRD evolution are found in the studies of \citet{Finkelstein+15}, \citet{McLeod+16} and \citeauthor{Bhatawdekar+18} (\citeyear{Bhatawdekar+18}; see also \citealt{Ellis+13} and \citealt{Madau+18}).
However, \citet{Oesch+18} claim that the SFRDs evolve more rapidly by $(1+z)^{-10.9}$ at $z>8$ that is shown with the blue solid curve in Figure \ref{fig:GSMD}.
There are two scenarios of the SFRD evolutionary trends, the smooth evolution of $(1+z)^{-3.9}$ and the rapid evolution of $(1+z)^{-10.9}$.
We calculate $\chi^2$ values with our GSMDs for the two functions of the smooth evolution and the rapid evolution.
The $\chi^2$ value is the smaller for the smooth evolution than for the rapid evolution by a factor of $\approx2.8$, implying that our results support the smooth evolution of the SFRDs. The smooth evolution of the SFRDs suggests that the star-formation rate per dark-matter mass increases at $z>8$ \citep{Harikane+18a, Oesch+18}.

\begin{deluxetable}{cc}
\tablecaption{Galaxy stellar mass densities.
\label{tab:GSMD}}
\tablehead{
\colhead{$z$}	&	\colhead{$\log\rho_\star~[\MS~\mathrm{Mpc}^{-3}]$}
}
\startdata
$6-7$	&	$6.21_{-0.37}^{+0.39}$	\\
$8$		&	$5.64_{-0.51}^{+0.52}$	\\
$9$		&	$5.91_{-0.65}^{+0.75}$	\\
\enddata
\end{deluxetable}

\subsection{Size-Mass Relations}

We compare the distribution of the local GCs and our dropouts in the parameter space of $R_\mathrm{e}$ and $M_\star$.
The GCs in the Milky Way today have $R_\mathrm{e}$ values up to $\approx40$ physical pc (e.g. Pal 5 and Pal 14) and $M_\star$ up to $\sim10^7~\MS$ (e.g. NGC 5139) that are shown in the catalog compiled by \citet{Baumgardt+18}.
We thus regard the dropouts with $R_\mathrm{e}\leq40$ physical pc and $M_\star\leq10^7~\MS$ as early GC candidates.

\begin{figure*}
\epsscale{1.2}
\plotone{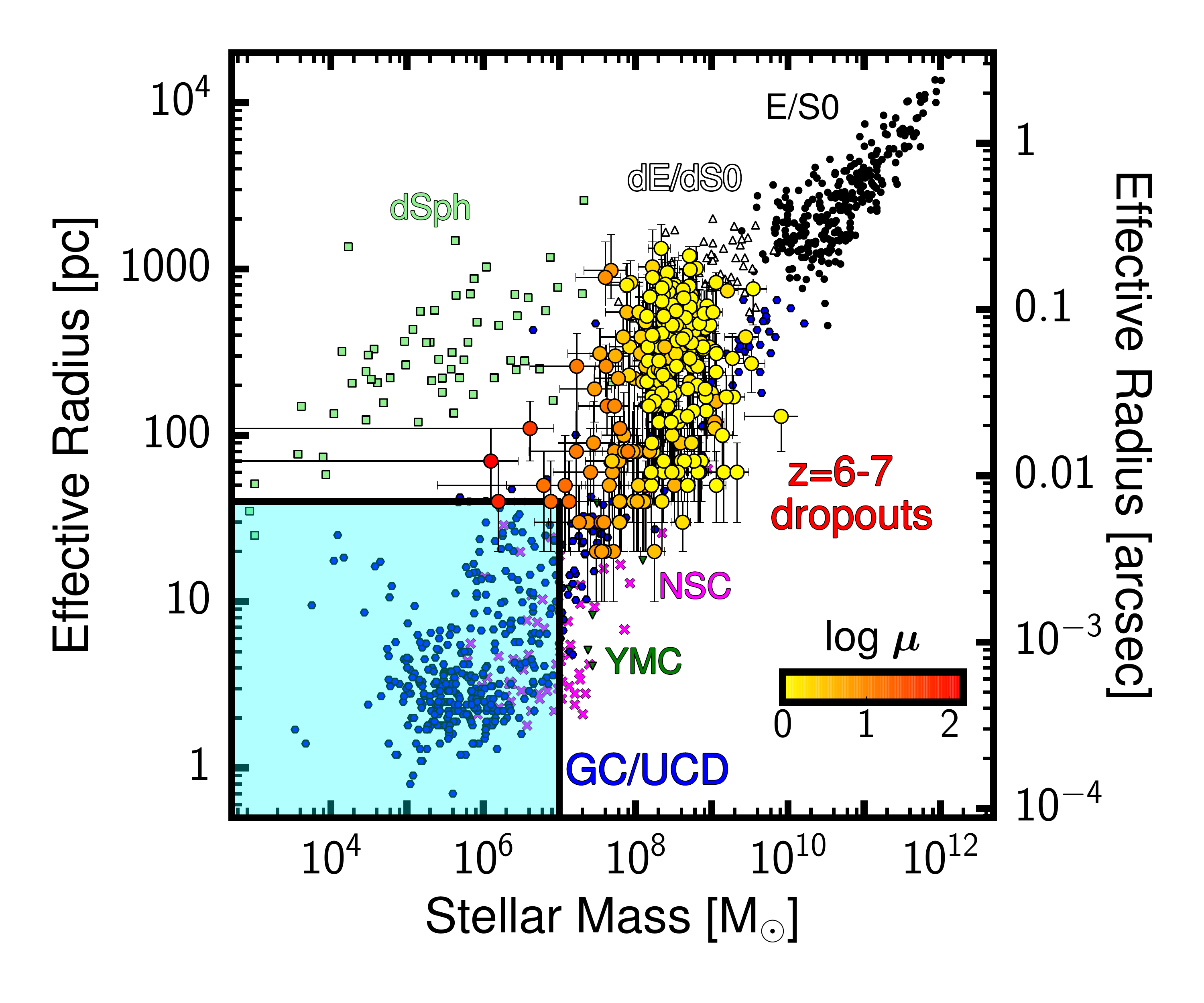}
\caption{Effective radius on the source plane, as a function of $M_\star$. Our $i$-dropouts are presented with the large circles whose colors indicate the magnification factor values that are defined with the color bar at the bottom right. The other symbols show the distributions of the local elliptical/S0 galaxies (E/S0s; black circles), dwarf elliptical/S0 galaxies (dEs/dS0s; gray triangles), dwarf spheroids (dSphs; light-green squares), nuclear star clusters (NSCs; orange crosses), young massive clusters (YMCs; green down-pointing triangles), and globular clusters/ultra-compact dwarfs (GCs/UCDs; blue hexagons) obtained by \citet{Norris+14} and the references therein. The cyan shade represents the region where the local GCs are located. We define the dropouts in this region as the early GC candidates. The right vertical axis shows $R_\mathrm{e}$ in the angular scale at $z=6$.}
\label{fig:SizeMass_z6-7}
\end{figure*}

\begin{figure}
\epsscale{1.2}
\plotone{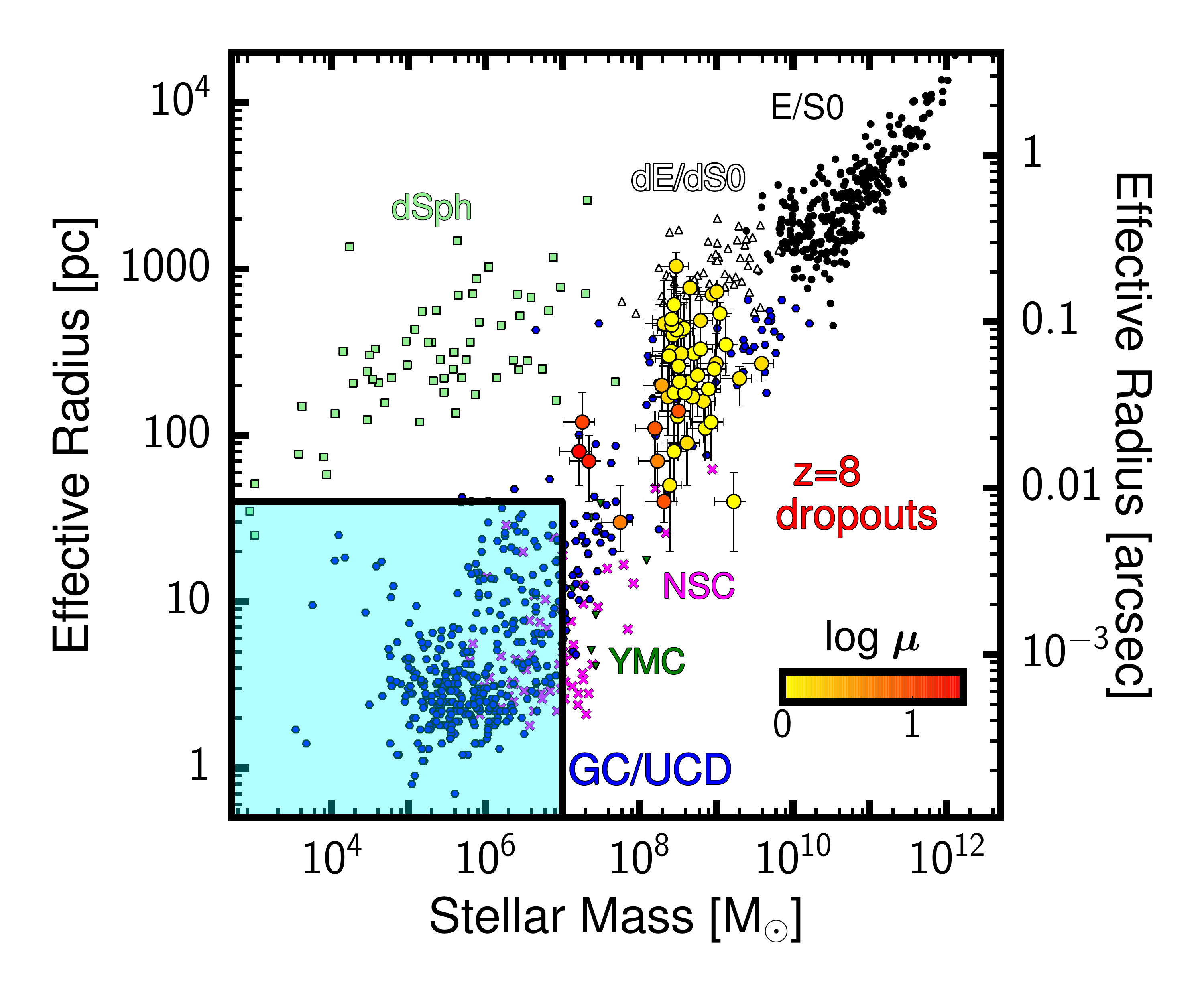}
\caption{Same as Figure \ref{fig:SizeMass_z6-7}, but for $Y$-dropouts. The right vertical axis shows $R_\mathrm{e}$ in the angular scale at $z=8$.}
\label{fig:SizeMass_z8}
\end{figure}

\begin{figure}
\epsscale{1.2}
\plotone{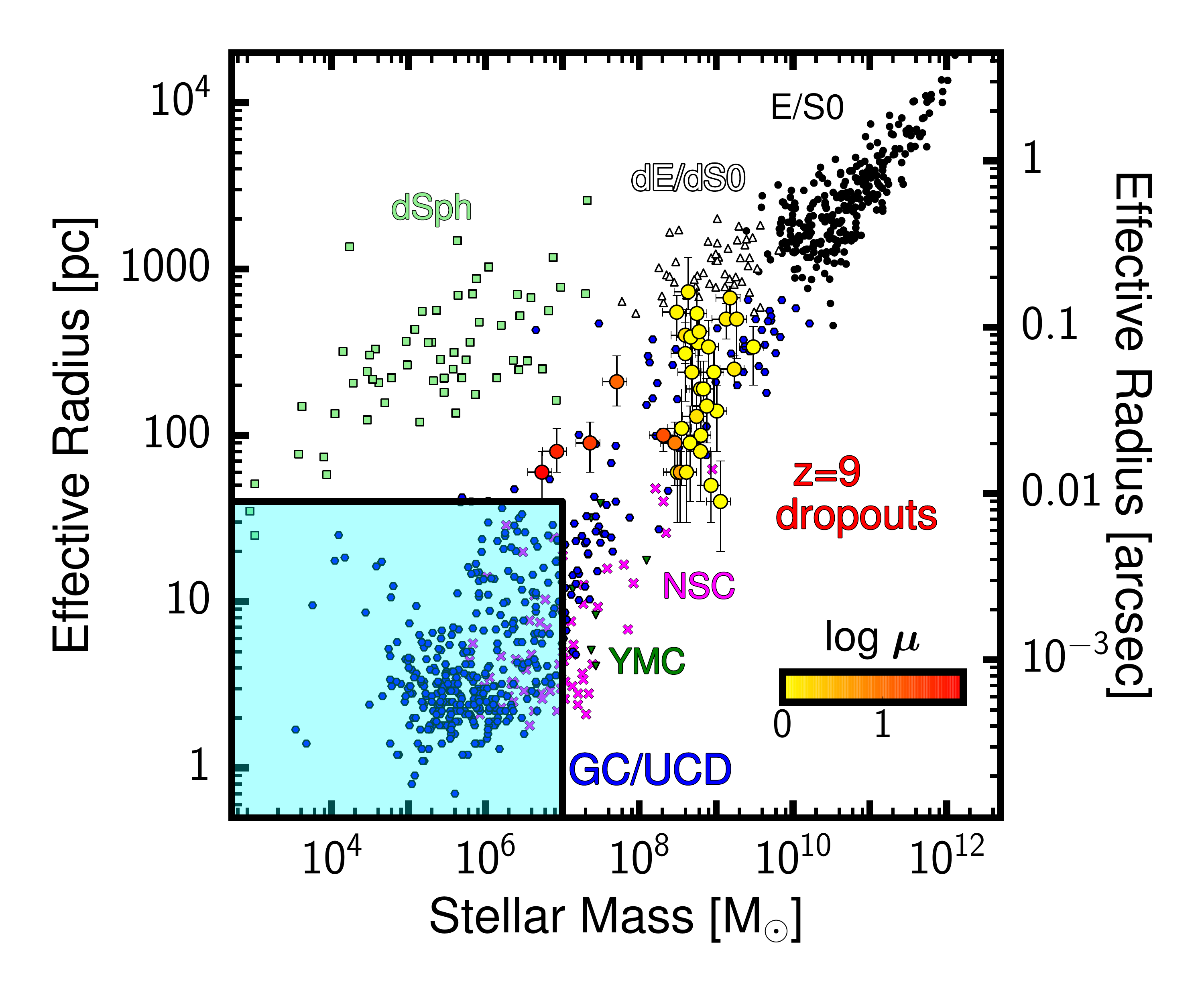}
\caption{Same as Figure \ref{fig:SizeMass_z6-7}, but for $J$-dropouts. The right vertical axis shows $R_\mathrm{e}$ in the angular scale at $z=9$.}
\label{fig:SizeMass_z9}
\end{figure}

We plot the $R_\mathrm{e}$-$M_\star$ values of our dropouts at $z\sim6-7$, $8$, and $9$ in Figure \ref{fig:SizeMass_z6-7}, \ref{fig:SizeMass_z8}, and \ref{fig:SizeMass_z9}, respectively.
The values of $R_\mathrm{e}$ are taken from \citet{Kawamata+18}, who obtain the best-fit value by the fit of ellipsoidal S\'{e}rsic profiles to the galaxy profiles that are observed with the lensing effects.
We measure $M_\UV$ of the dropouts in the same manner as those for the stacked images.
The values of $M_\UV$ are converted to $M_\star$ with the best-fit $M_\star$-$M_\UV$ relations obtained in Section \ref{subsec:masslumi}.

We compare the $R_\mathrm{e}$-$M_\star$ distributions with those of the stellar systems in the local universe (\citealt{Norris+14} and the references therein).
We find that two of our dropouts, HFF5C-4260-1364 and HFF5C-4039-1566, \footnote{These IDs are identical to the ones by \citet{Kawamata+18} and \citet{Ishigaki+18}.}
meet the criteria of the early GC candidates.
The parameters of the candidates are listed in Table \ref{tab:EGCC}.

As stated in the introduction, \citet{Bouwens+17} study the $R_\mathrm{e}$-$M_\star$ distributions of the $z\sim6-8$ dropouts.
However, \citet{Bouwens+17} use only the first four HFF cluster (A2744, M0416, M0717, and M1149) data that were available at the time of their publication. No studies of $R_\mathrm{e}$-$M_\star$ distribution for $z\gtrsim 6$ are conducted in the rest of two clusters, A1063 and A370. Our early GC candidates of HFF5C-4260-1364 and HFF5C-4039-1566 are placed in the A1063 field. These early GC candidates are identified for the first time.

In \citet{Bouwens+17}, there are 18 $z\sim6-8$ dropouts meeting the criteria of the early GC candidates with their stellar masses obtained with the assumption of the 100-Myr star-formation duration time (Equation \ref{eqn:ml_B17_100My}).
One out of the 18 dropouts, M0416I-6115434445, is also selected in our study that refers to the dropout as HFF2C-1156-3446.
This dropout is also reported as a compact star-forming region GC1 in \citet{Vanzella+17a}.
The stellar masses and effective radii obtained in our study, \citet{Bouwens+17}, and \citet{Vanzella+17a}
are broadly consistent within the $1-2\sigma$ uncertainties.
Note that another dropout from \citet{Bouwens+17}, M0416I-6118103480, is selected as the dropout HFF2C-1181-3480 in the original sample of \citet{Kawamata+18} and \citet{Ishigaki+18}, but removed from our sample in Section \ref{subsec:hst}, due to the possible detection of the blue continuum flux. Because our detection threshold would be conservative, the rest of the 16 dropouts shown in \citet{Bouwens+17} are not securely detected in our analysis.

The previous studies suggest that the GCs in the Milky Way form almost instantaneously at $z\gtrsim5$, arguing that the stars inside the GCs have high number densities and tight isochrones on the HR diagram \citep[e.g.,][]{VandenBerg+96, Forbes+10}.
The dropouts that we select are thus candidates of star clusters that should be a part or a dominant component of high-redshift low-mass galaxies, some of which may be related to GCs today.
Our calculations for observational feasibilities suggest that these candidates can be spectroscopically confirmed with the {\it James Webb} Space Telescope and the Extremely Large Telescopes such as the Thirty Meter Telescope.

\begin{deluxetable*}{ccccccc}
\tablecaption{Catalog of the early GC candidates.
\label{tab:EGCC}}
\tablehead{
\colhead{ID} & \colhead{R.A. (J2000)} & \colhead{decl. (J2000)} & \colhead{$M_\UV$ [mag]} & \colhead{$M_\star~[10^6~\MS]$} & \colhead{$R_\mathrm{e}$ [physical pc]} & \colhead{$\mu$}
}
\startdata
HFF5C-4260-1364	&	$342.177526$	&	$-44.526786$	&	$-15.7\pm0.2$		&	$7.7\pm7.1$	&	$40_{-20}^{+30}$	&	19.12	\\
HFF5C-4039-1566	&	$342.168321$	&	$-44.532412$	&	$-14.2\pm0.1$		&	$1.6\pm2.0$	&	$40_{-20}^{+30}$	&	77.69	\\
\enddata
\end{deluxetable*}

\section{Summary} \label{sec:summary}

We present stellar populations of low-mass dropout galaxies at $z\sim 6-9$ down to a stellar mass of $M_\star\sim10^6~\MS$ identified with the full data sets of the HFF. The stellar populations are studied with the HST and {\it Spitzer} deep optical to NIR images and the cluster-lensing magnification effects. The major results of our study are summarized below.

\begin{enumerate}
\item{We investigate the stellar populations of the dropout galaxies with the optical/NIR photometric measurements and the {\tt BEAGLE} SED modeling tool. We estimate $M_\star$ as a function of the UV magnitude $M_\UV$ at $M_\star\sim10^6-10^9~\MS$. We find that our best-estimate $M_\star/L_\UV$ is consistent with the constant $M_\star/L_\UV$ within the $1\sigma$ uncertainties.
Our best-estimate $M_\star/L_{\rm UV}$ function is comparable to a model of star-formation duration time of 100 Myr, but $2-7$ times higher than the one of 10 Myr assumed in a previous study at the $5\sigma$ level.}
\item{We apply the best-estimate $M_\star$-$M_\UV$ relation (corresponding to $M_\star/L_\UV$) to the LFs of \citet{Ishigaki+18} to derive the GSMFs at $z\sim6-9$. The GSMFs agree with those obtained by previous studies broadly at $M_\star\gtrsim10^8~\MS$, extending to the low-mass limit of $\sim10^6~\MS$. We reach this low-mass limit, exploiting the deep HFF full data sets and the gravitational-lensing magnification effects.}
\item{Integrating the GSMFs, we find that the GSMDs $\rho_\star$ smoothly increases from $\log(\rho_\star/[\MS~\mathrm{Mpc}^{-3}])=5.91_{-0.65}^{+0.75}$ at $z\sim9$ to $6.21_{-0.37}^{+0.39}$ at $z\sim6-7$. This trend is consistent with the one estimated from time integration of the SFRD evolution compiled by \citet{Madau+14}.}
\item{The stellar masses of the dropouts are presented as a function of effective radius $R_\mathrm{e}$ with our best-fit $M_\star$-$M_\UV$ relations. We find that the two dropouts have low stellar masses ($M_\star\leq10^7~\MS$) and compact morphologies ($R_\mathrm{e}\leq40$ physical pc), which are comparable with those of GCs in the Milky Way today. These dropouts are candidates of star clusters that should be a part or a dominant component of high-redshift low-mass galaxies. These dropouts may be related to the present GCs, given the fact that the old stellar populations are found in GCs of the Milky Way today.}
\end{enumerate}

We are grateful to Michiko Fujii, Seiji Fujimoto, Ryo Higuchi, Ryohei Itoh, Nobunari Kashikawa, Shiro Mukae, and Haibin Zhang for useful comments and discussions.
We particularly thank Masafumi Ishigaki and Ryota Kawamata for their sample catalogs and mass models.
GB acknowledges financial support through PAPIIT projects IG100319 from DGAPA-UNAM.
This work is supported by World Premier International Research Center Initiative (WPI Initiative), MEXT, Japan, and KAKENHI (15H02064, 17H01110, 17H01114, 15H05892, and 18K03693) Grant-in-Aid for Scientific Research (A) through Japan Society for the Promotion of Science.
This work is based on data and catalog products from HFF-DeepSpace, funded by the National Science Foundation and Space Telescope Science Institute (operated by the Association of Universities for Research in Astronomy, Inc., under NASA contract NAS5-26555).


\bibliographystyle{apj}
\bibliography{cite}


\end{document}